\documentclass[aps,prd,twocolumn,showpacs,superscriptaddress,nofootinbib]{revtex4-2}

\usepackage{amsmath,amssymb}
\usepackage{graphicx}
\usepackage{hyperref}
\usepackage{orcidlink}
\usepackage{gensymb}
\usepackage[normalem]{ulem}

\usepackage{array}
\newcolumntype{R}[1]{>{\raggedleft\arraybackslash}p{#1}} % right-aligned

\begin{document}

\title{A SHIFT of Perspective: Observing Neutrinos at CMS and ATLAS}

\author{Alfonso Garcia-Soto \orcidlink{0000-0002-8186-2459}}
\affiliation{Instituto de Física Corpuscular (IFIC), CSIC‐UV, 46980 Paterna, València, Spain}

\author{Jeremi Niedziela \orcidlink{0000-0002-9514-0799}}
% \affiliation{Deutsches Elektronen-Synchrotron DESY, Notkestr. 85, 22607 Hamburg, Germany}
\affiliation{Interuniversity Institute for High Energies (IIHE), Vrije Universiteit Brussel, Pleinlaan 2, 1050 Brussels, Belgium}

\begin{abstract}

The SHIFT@LHC proposal introduced a novel shifted gaseous fixed-target concept at the LHC to search for exotic particles. In this letter, we explore an entirely different physics opportunity enabled by this setup: the observation of neutrinos in general-purpose LHC detectors.
Using simulations of proton–gas collisions, hadron propagation, and neutrino interactions, we estimate that $\mathcal{O}(10^4)$ muon-neutrino and $\mathcal{O}(10^3)$ electron-neutrino interactions, with energies from 20~GeV to 1~TeV, would occur in the CMS and ATLAS detectors with 1\% of the LHC Run-4 integrated luminosity ($\approx4\cdot10^{25}$ protons-on-target).
This unique configuration provides access to hadron production in the pseudorapidity range $5<\eta<8$, complementary to existing LHC detectors.
If realized, this would mark the first detection of neutrinos in a hadron collider detector, demonstrating the feasibility of such measurements in this experimental environment.
\end{abstract}

\maketitle

\section{Introduction}

Since the 1980s, it has been recognized that hadron colliders operating in the TeV energy range naturally provide an intense and highly collimated flux of forward neutrinos~\cite{DeRujula:1984pg}.
Recently, the FASER~\cite{FASER:2023zcr} and SND@LHC~\cite{SNDLHC:2024qqb} experiments reported the first observation of such neutrinos from proton–proton collisions at the LHC.
This milestone has opened a rich physics program, including studies of hadron production~\cite{Kling:2021gos,Francener:2025pnr}, measurements of neutrino cross sections~\cite{FASER:2024hoe,FASER:2024ref}, and searches for new physics~\cite{FASER:2023tle,FASER:2024bbl}.
Several new detector concepts have already been proposed, suggesting that up to a million neutrino interactions, with energies peaking at a few TeV, could be observed during the high-luminosity run of LHC~\cite{Feng:2022inv,Kamp:2025phs,Ariga:2025gtj,MammenAbraham:2025gai}.

Other strategies for detecting LHC neutrinos have also been considered.
For instance, first estimates indicate that a handful of neutrinos from W boson decays might be observed in the CMS high-granularity calorimeter during the high-luminosity run of the LHC~\cite{Foldenauer:2021gkm}.
Another idea is to place a detector near the LHC beam dump, where roughly 20 neutrinos with energies above 10 GeV would interact every time a beam is aborted~\cite{Kelly:2021jgj}. There have also been proposals to detect atmospheric and astrophysical neutrinos with ATLAS~\cite{Kopp:2007ai,Wen:2023ijf,Ghosh:2024ryg}.

A recent proposal, SHIFT@LHC~\cite{Niedziela:2024khw}, envisions installing a gaseous fixed target---similar to SMOG/SMOG2~\cite{Hadjidakis:2018ifr,Franzoso:2022tfv}---in the LHC tunnel at an $\mathcal{O}(100~\rm m)$ distance from the main interaction points.
The fixed target collisions would happen concurrently with the standard collider-mode events, potentially overlapping with the O(100) simultaneous collisions in ATLAS/CMS. However, even in such a scenario, neutrino interactions can be easily disentangled from the primary event based on the jet and lepton pointing angles, displacement, and the formation of a common secondary vertex.
While the original study focused on searches for long-lived particles, here we explore a fundamentally different physics opportunity: using SHIFT as a source of neutrinos that can be detected in ATLAS and CMS.
In this letter, we show that this configuration would generate an intense flux of neutrinos with energies ranging from a few GeV to the TeV scale, leading to a sizable number of interactions in the ATLAS~\cite{ATLAS:2008xda} and CMS~\cite{CMS:2008xjf} detectors.
This opens the possibility of performing for the first time neutrino measurements directly within general-purpose LHC detectors, thereby extending the physics reach of the original concept.

This energy range has previously been explored with dedicated neutrino beams at the SPS and Tevatron by detectors such as CCFR~\cite{Sakumoto:1990py}, NOMAD~\cite{NOMAD:1997pcg}, CHORUS~\cite{CHORUS:1997wxi}, NuTeV~\cite{NuTeV:1999uck}, and MINOS~\cite{MINOS:1995txm}. 
In this work, we highlight differences with respect to those experiments, which could open the possibility to address several physics questions. 
In particular, we discuss aspects that would provide valuable input for experiments sensitive to GeV-scale atmospheric neutrinos such as KM3NeT-ORCA~\cite{KM3Net:2016zxf}, IceCube-DeepCore/Upgrade~\cite{IceCube:2011ucd,IceCube-PINGU:2014okk}, DUNE~\cite{DUNE:2020lwj}, and Hyper-Kamiokande~\cite{Abe:2011ts}.

\section{Methodology}\label{sec:methodology}

In proton–nucleus collisions, the dominant source of neutrinos originates from Quantum Chromodynamics (QCD) processes mediated by quarks and gluons.
These partons hadronize, and the resulting hadrons decay, producing neutrinos of various flavors.
We simulated these processes using \texttt{PYTHIA8}~\cite{Sj_strand_2015}, with a 6.8 TeV proton beam impinging on a stationary proton target.
Default settings were applied, and events were generated in bins of the hard-process transverse momentum ($\hat{p}_{T}$) to obtain a realistic sample for further study.

The proposed fixed-target collision point is located approximately 160 m upstream of the CMS or ATLAS interaction point, within the Long Straight Section of the LHC~\cite{Bruning:2004ej}.
This region is relatively free of massive structures: although there is material associated with crab cavities, warm magnets, support structures, pipes, and cables, there are no cryostats.
Moreover, most collision products are produced at angles of $\approx1\degree$, allowing them to avoid the surrounding instrumentation to a large degree. Consequently, the dominant material relevant for secondary interactions is the rock surrounding the LHC tunnel. Nevertheless, a detailed study with a proper simulation of the LHC instrumentation is needed to assess losses due to the absorption. These will reduce the total neutrino flux, however other improvements (such as better reconstruction algorithms) may effectively balance these effects out.

\begin{figure*}[t!]
\centering
\includegraphics[width=1.0\textwidth]{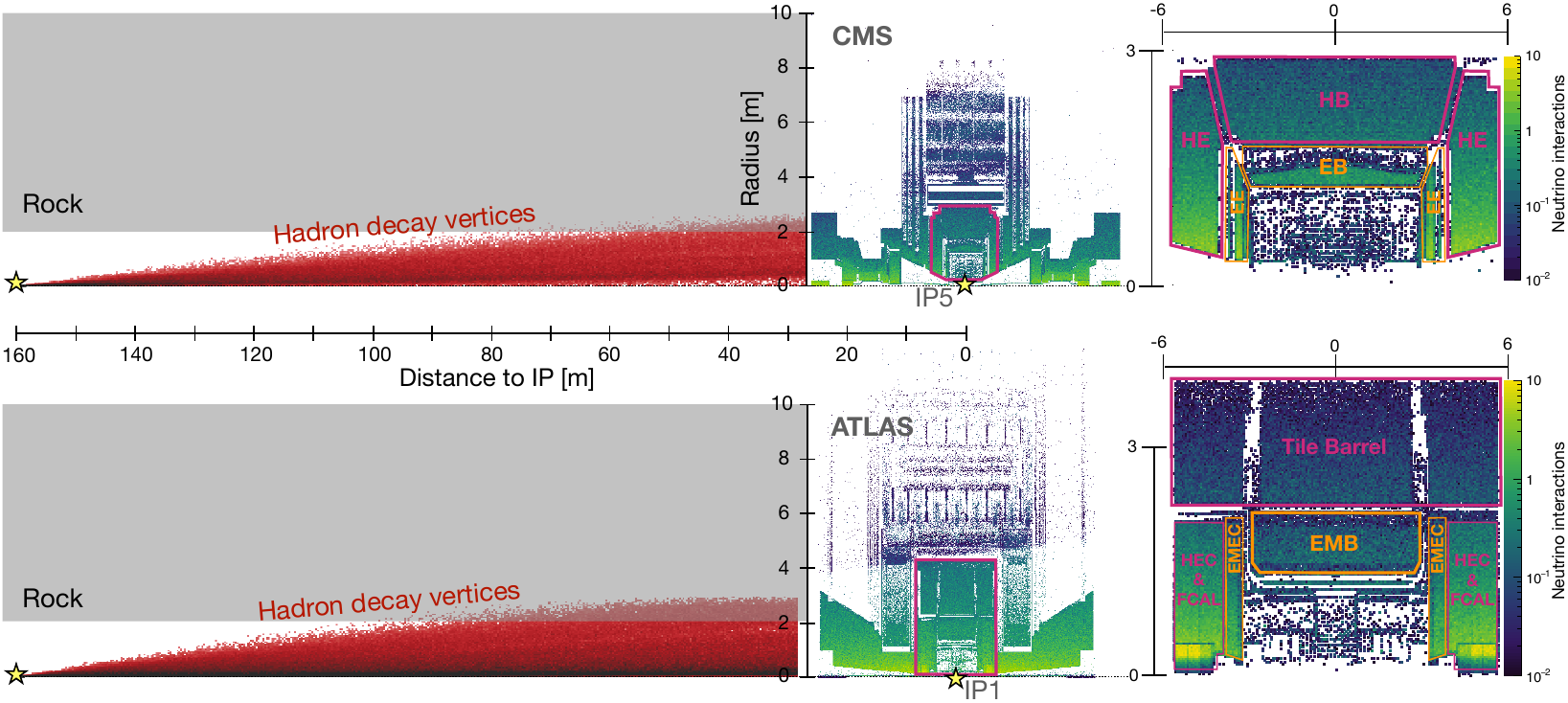}
\caption{\label{fig:locations} Distribution of the production and interaction vertices of neutrinos with the gaseous target placed 160 meters away from IP5 (top) and IP1 (bottom). The production points are shown for neutrinos interacting in the CMS and ATLAS calorimeters, in which darker (lighter) red indicates the regions where there are more (fewer) decay vertices. The interaction vertices are shown for all subdetectors, in which yellow (blue) colors indicate the regions where there are more (fewer) interactions. The fiducial volumes of the calorimeters used in this analysis are marked with red lines.
The right-hand panels provide zoomed views of the calorimeters in both detectors, showing the spatial distribution of neutrino interactions with a $200\times200$ binning.}
\end{figure*}

We consider three main scenarios: (a) a neutrino is produced before the mother particle reaches the rock; (b) a neutrino originates from a hadron decay within the rock; and (c) a neutrino originates from a muon decay inside the rock. 
In case (a), no suppression from the material is applied, since the probability of a neutrino stopping within a few tens of meters of rock is negligible.
For case (b), we performed a \texttt{GEANT4}~\cite{GEANT4:2002zbu} simulation of standard rock to evaluate the impact of the material on hadrons and muons. Hadrons interact frequently with the rock and are all stopped within less than 1 m, irrespective of their type or energy. Therefore, neutrinos whose mother hadron travels more than 1 m inside the rock are discarded.
Finally, for case (c), we found that no muons decayed to neutrinos within our decay volume, due to their relatively long lifetime.

The propagation of neutrinos and their subsequent interactions in CMS and ATLAS\footnote{CMS and ATLAS geometries for Phase-1 are obtained from \url{https://root.cern.ch/files/}. The impact of the slightly different Phase-2 geometries on neutrino measurements is expected to be minor.} are simulated with \texttt{GENIE}~\cite{Andreopoulos:2009rq}.
We focus on charged-current (CC) interactions above 20 GeV, for which only deep inelastic scattering is relevant.
For the modeling of neutrino–nucleus interactions, we use the tune \texttt{G18\_02a}~\cite{GENIE:2021zuu} for energies from 20 GeV to 1 TeV, and \texttt{GHE19\_00b}~\cite{Garcia:2020jwr} for energies above 1 TeV. 
The former employs the Bodek–Yang prescription for nuclear structure functions~\cite{Bodek:2004pc}, while the latter relies on the CSMS model~\cite{Cooper-Sarkar:2011jtt}.

Our main results focus on neutrino interactions within the hadronic and electromagnetic calorimeters of CMS~\cite{CERN-LHCC-97-031,CERN-LHCC-97-033} and ATLAS~\cite{CERN-LHCC-96-042,CERN-LHCC-96-041}, which are the innermost calorimetric systems and are surrounded by the muon detection layers~\cite{CERN-LHCC-97-022,Layter:343814}. 
The outer muon system can be used to tag muons from hadronic decays, enabling the isolation of neutrino interactions inside the calorimeters. 
We require the outgoing charged lepton and hadronic shower to carry an energy above 3 and 10 GeV, respectively. Particles with these energies would produce a detectable signal in the detectors~\cite{ATLAS:2019qmc,ATLAS:2020auj,CMS:2024qyo,Elmetenawee:2023xyl}, which can be used to reconstruct the neutrino vertex. 
The electrons, muons, and hadronic jets can be distinguished from each other almost perfectly, with a typical mistag rate below a per mille level. 
We recognize that reconstructing a displaced jet+lepton vertex and designing suitable triggers is challenging. Nevertheless, for this study, we assume that with the advances in computational methods and given the high granularity and the timing capabilities of the LHC detectors, these difficulties will be overcome, and highly efficient reconstruction algorithms will be developed.

The distribution of neutrino production and interaction vertices from the full simulation chain is shown in Fig.~\ref{fig:locations}. 
Red points indicate the distribution of hadron decay vertices that produce neutrinos subsequently interacting in the calorimeters.
The decays of pions and kaons producing neutrinos decrease exponentially along the beam axis, whereas prompt neutrino production is strongly peaked near the gaseous target.
The distribution of neutrino interaction vertices within CMS and ATLAS also shows a clear radial dependence related to the forward nature of the hadronic showers.

\section{Results}

Table~\ref{tab:rates} summarizes the expected number of neutrino interactions in the calorimeters of CMS and ATLAS, assuming that 1\% of the Run-4 integrated luminosity is dedicated to this study, corresponding to 7.15~fb$^{-1}$~\cite{Tomas:2022mzg}, and equivalent to $\approx4\cdot10^{25}$ protons-on-target.
Results are separated by different neutrino flavors.

\begin{table}[h!]
\centering
\begin{tabular}{lr|R{0.8cm}R{0.8cm}R{0.8cm}R{0.8cm}R{0.8cm}R{0.8cm}}
CMS & $A$ & $\nu_\mu$ & $\bar{\nu}_\mu$ & $\nu_e$ & $\bar{\nu}_e$ & $\nu_\tau$ & $\bar{\nu}_\tau$ \\
\hline
HE &  64 & 3951 & 1230 & 220 &  71 & 3 & 1 \\
HB &  64 & 1245 &  320 &  62 &  16 & 1 & $-$ \\
EE & 171 &  679 &  212 &  38 &  12 & 1 & $-$ \\
EB & 171 &  494 &  139 &  27 &   7 & $-$ & $-$ \\
\hline
\hline
ATLAS & A & $\nu_\mu$ & $\bar{\nu}_\mu$ & $\nu_e$ & $\bar{\nu}_e$ & $\nu_\tau$ & $\bar{\nu}_\tau$ \\
\hline
FCAL        & 170/63 &  9584 &  3290 & 573 & 213 & 8 & 4 \\
HEC         &     63 &  4187 &  1311 & 238 &  79 & 2 & 1 \\
Tile Barrel &     56 &  1085 &   267 &  52 &  12 & 1 & $-$ \\
EMEC        & 207/56 &   699 &   210 &  40 &  13 & 1 & $-$ \\
EMB         & 207/56 &   411 &   110 &  21 &   6 & $-$ & $-$ \\
\end{tabular}
\caption{\label{tab:rates} Expected number of CC neutrino interactions in the different calorimeters of CMS and ATLAS (see Fig.~\ref{fig:locations}). The second column shows the atomic mass number $A$ of the main materials in each module.}
\end{table}

\begin{figure}[t!]
\centering
\includegraphics[width=.45\textwidth]{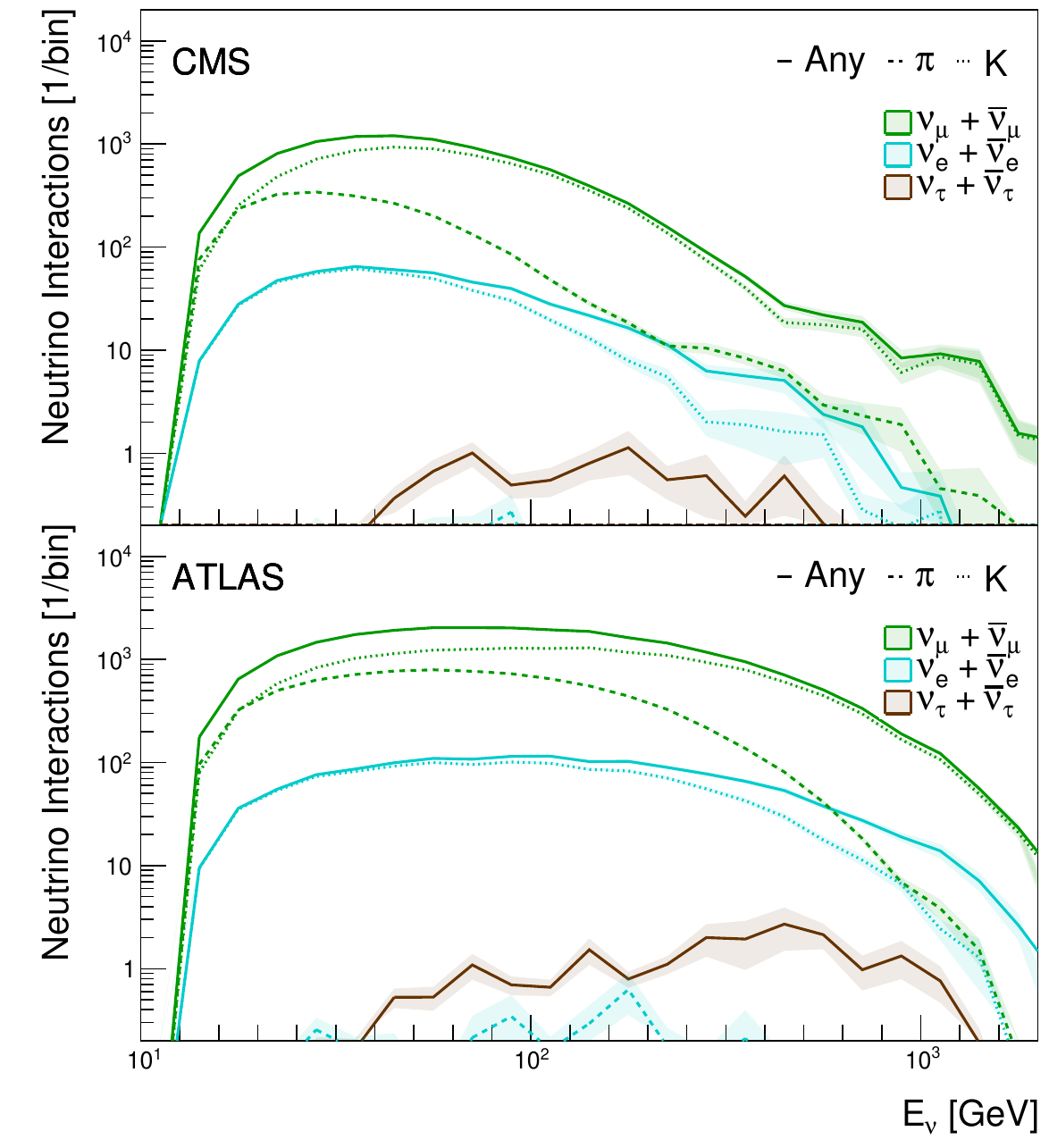}
\caption{\label{fig:flavors} Energy distribution of neutrino and antineutrino CC interactions in the calorimeters of CMS (top) and ATLAS (bottom). Colors represent different neutrino flavors. Dashed and dotted lines indicate neutrinos produced in pion and kaon decay, respectively.}
\end{figure}

Most of the interactions are $\nu_\mu$ CC events, spanning from 20 GeV up to a few TeV, with a peak between 20–100 GeV, as shown in Fig.~\ref{fig:flavors}. 
The dominant contribution arises from kaon decays, whereas pion decays become relevant below 15 GeV.
This behavior reflects the shorter decay length of charged kaons with respect to pions, which enhances their probability to decay within the available baseline.
In CMS, more than 70\% of all interactions occur in the hadronic calorimeters, whose primary absorber is brass (both the endcap and the barrel). 
Nevertheless, the electromagnetic calorimeters, made out of PbWO$_{4}$, will also contain a non-negligible fraction of events.
In ATLAS, the majority of interactions will occur in the forward calorimeter (whose absorber is a combination of tungsten and copper), followed by the hadronic end-cap (copper), and the hadronic barrel (steel). 

\begin{figure}[t!]
\centering
\includegraphics[width=.45\textwidth]{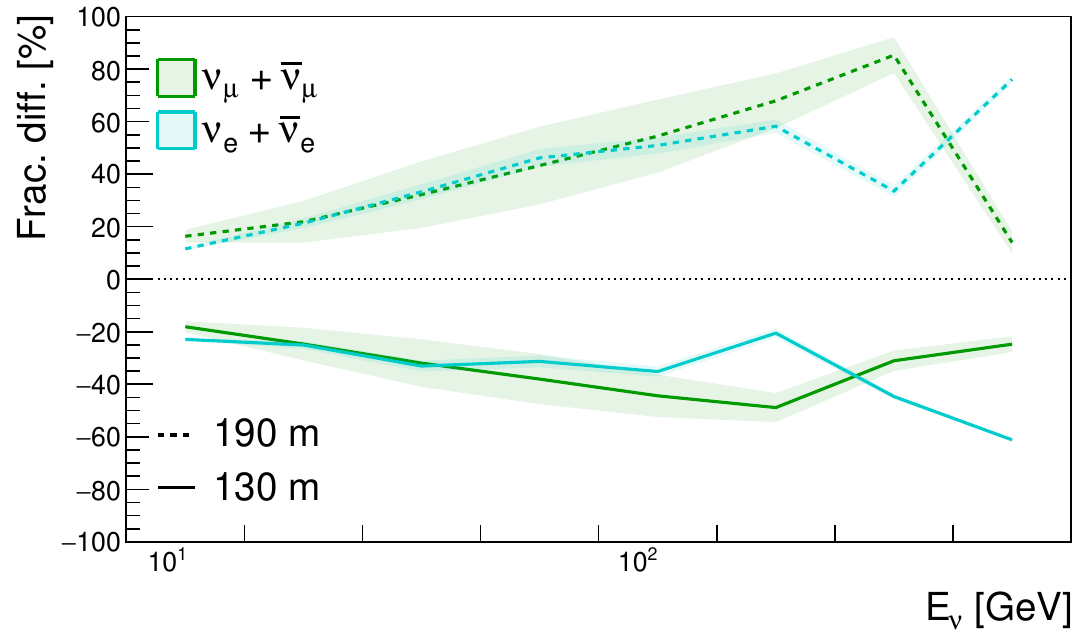}
\caption{\label{fig:distance} Relative change in the electron- and muon-neutrino interaction rate as a function of the target location with respect to IP5.}
\end{figure}

As described in Sec.~\ref{sec:methodology}, our nominal assumption is a fixed-target located 160 m upstream of IP5 or IP1. This site was chosen for its minimal surrounding instrumentation and for maximizing sensitivity to certain new-physics scenarios.
To study the dependence of the target placement on the neutrino yield, we varied the distance by $\pm 30$ meters with respect to this nominal position in IP5. 
The results, shown in Fig.~\ref{fig:distance}, indicate that the rates increase (decrease) by about 30\% when the target is placed 30 m farther (closer) from IP5 .
This behavior is explained by the longer decay path available to hadrons at larger distances.
A more detailed simulation of the LHC tunnel geometry would be required to determine the optimal target location for neutrino studies.

\begin{figure}[t!]
\centering
\includegraphics[width=.45\textwidth]{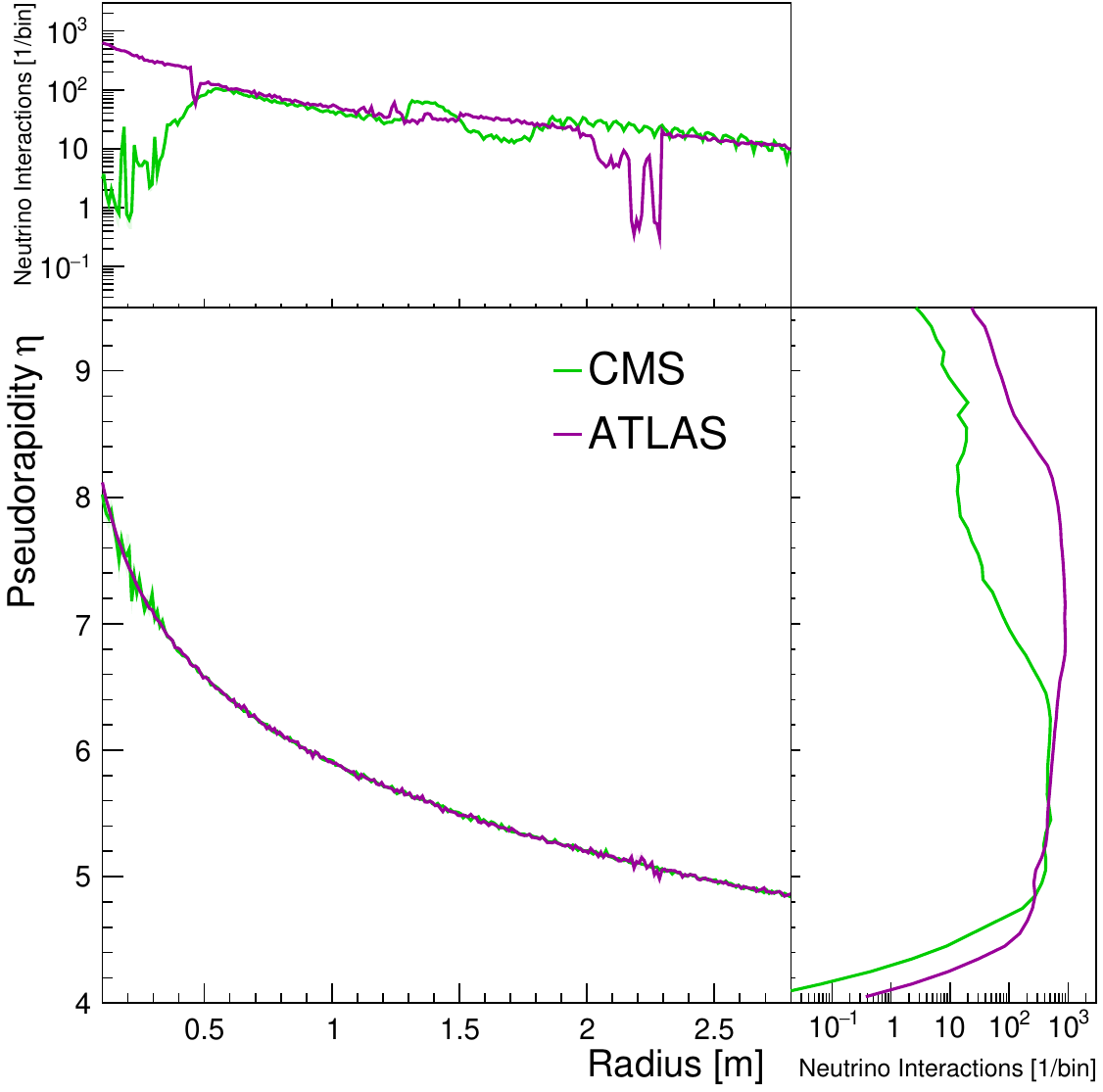}
\caption{\label{fig:pseudo} Parent pseudorapidity and radial distributions of $\nu_\mu+\bar{\nu}_\mu$ CC interactions in CMS and ATLAS. 
The center panel shows the average pseudorapidity as a function of the radial distribution.}
\end{figure}

This detector configuration also allows access to different pseudorapidity regions, depending on the position of the neutrino interaction vertex within the calorimeter. 
Figure~\ref{fig:pseudo} shows the average pseudorapidity of $\nu_\mu$ CC interactions as a function of the radial distance of the vertex from the beam axis.
It can be observed that ATLAS could probe larger regions of pseudorapidity because its end-caps calorimeters have more coverage in the forward direction than CMS.

\begin{figure}[t!]
\centering
\includegraphics[width=.45\textwidth]{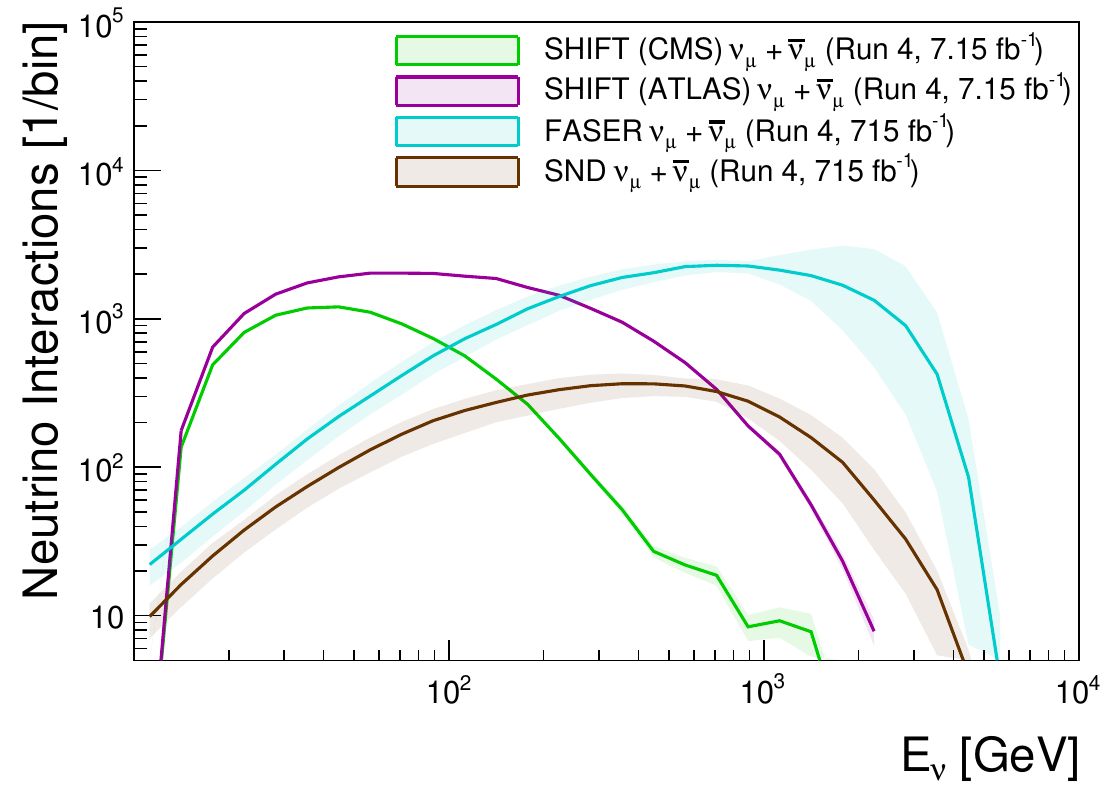}
\caption{\label{fig:energy} Expected energy distribution of muon neutrinos measurable in CMS and ATLAS with SHIFT, compared to FASER and SND@LHC projections \cite{Kling:2021gos}. FASER and SND@LHC results are scaled to 715~fb$^{-1}$, corresponding to the LHC Run-4 integrated luminosity. For SHIFT, 1\% of the Run-4 integrated luminosity is assumed, following \cite{Niedziela:2024khw}.}
\end{figure}

Finally, in Fig.~\ref{fig:energy}, we show how SHIFT allows us to probe the LHC neutrinos at the energy range complementary to what FASER and SND@LHC measure, with the potential of providing crucial input for the atmospheric neutrino experiments.

\section{Physics opportunities}

This preliminary study highlights several physics opportunities that could be explored with the proposed fixed-target configuration of SHIFT.

Placing a gaseous fixed target 160 m upstream of CMS or ATLAS provides access to pion and kaon production in the pseudorapidity range $4<\eta<9$, with maximal sensitivity between $5<\eta<6.5$ in CMS and $5<\eta<8$ in ATLAS. 
This region is inaccessible in nominal pp collisions at the LHC: neither CMS nor ATLAS can probe such forward kinematics, nor can very-forward detectors such as FASER or SND@LHC.
Other setups, including TOTEM~\cite{CMS:2014kix} or the SMOG program at LHCb, cover parts of this pseudorapidity interval. 
However, the extended decay length available in the SHIFT configuration enables complementary studies of hadron propagation and decay before interaction, providing sensitivity to the relative pion and kaon contributions through the resulting neutrino sample.
In fact, our design offers the unique advantage of providing a direct proxy for the pseudorapidity of parent hadrons, which can be inferred from the location of the neutrino interaction vertex in the calorimeters, as shown in Fig.~\ref{fig:pseudo}.

In the energy range from a few tens to several hundreds of GeV, atmospheric neutrino flux uncertainties are at the level of $10\%$~\cite{Fedynitch:2022vty}, and are dominated by modeling of forward hadronic interactions.
The expected event rates in Table~\ref{tab:rates} indicate that, even assuming reconstruction efficiencies of order 10\%, hundreds to thousands of CC events could be measured depending on flavor and topology.
This corresponds to statistical uncertainties at the few-percent level on inclusive interaction rates.
Measurements of ratios such as $\bar{\nu}_{\mu}/\nu_{\mu}$ and $\nu_{e}/\nu_{\mu}$ could therefore reach statistical precisions comparable to current flux uncertainties, providing a direct experimental constraint on hadronic production in the forward region.
Furthermore, the possibility of injecting different gases into the target would allow dedicated studies of nuclear effects in hadron production. 
This is particularly relevant for atmospheric neutrinos, where the dominant production channels involve proton scattering on nitrogen and oxygen.

The proposed setup shares similarities with past fixed-target neutrino experiments such as MINOS~\cite{MINOS:2009ugl}, NOMAD~\cite{NOMAD:2007krq}, CHORUS~\cite{CHORUS:2005cpn}, and NuTeV~\cite{NuTeV:2005wsg}, which were sensitive to comparable neutrino energies and also used magnetized detectors. 
There are a few key differences in the neutrino beams: in our case, the protons are accelerated to higher energies, the target is a low-pressure gas rather than a dense medium, and no magnetic horns are used to focus hadrons. 
As a result, those past beams achieved much higher fluxes in the forward direction, enabling large samples of neutrino interactions, which cannot be reached with the proposed setup with limited luminosity.
As an example, the NuTeV experiment selected the order of one neutrino interaction per $10^{12}$ protons-on-target, whereas the configuration considered here would require $10^{21}$.
Therefore, the proposed setup cannot compete with those experiments in high-precision measurements of absolute $\nu_\mu$ CC cross sections.

Nevertheless, the CMS and ATLAS calorimeters have absorber materials of different nuclear compositions, allowing ratio measurements of neutrino cross sections on various nuclei.
Dedicated measurements in this energy regime remain scarce; among accelerator experiments, only MINERvA~\cite{MINERvA:2016oql} and CHORUS~\cite{CHORUS:2003qcm} have reported such studies using targets such as plastic, marble, iron, or lead. 
In the case of CHORUS, the available samples were of comparable size to those reported in Table~\ref{tab:rates}, achieving precisions of order 3\% on cross-section ratios.
The SHIFT configuration could reach comparable statistical precision for neutrinos and extend these studies to antineutrinos, for which there are no measurements.
In addition, interactions on materials such as tungsten or copper present in the calorimeters would provide new data points.
Finally, a particularly interesting aspect is the expected sample of $\nu_e+\bar{\nu}_e$ interactions, since $\nu_e$ cross-section measurements above 10 GeV remain very limited. 
With hundreds of reconstructed events, statistical uncertainties at the level of $10\%$ on flavor-dependent cross sections appear achievable. 

Finally, in this work, we have focused on event topologies that are most straightforward to reconstruct with CMS and ATLAS.
Future studies could extend the analysis to lower-energy events, NC interactions, and interactions occurring in the outer detector layers.
More detailed simulations will also be required to assess the effect of the surrounding material on hadron propagation from the target to the calorimeters.

\section{Challenges}

An important challenge of this setup will be the separation of neutrino interactions from the background of muons traversing the detector, which are produced in the same hadronic decays.

From the simulation, we observe that in the nominal configuration, with the target located 160 m from the CMS interaction point, if a neutrino is produced in the tunnel and subsequently interacts in the calorimeters, an average of around 5 muons are also produced in the tunnel.
Using the simulated production points and directions, we perform a ray-tracing propagation to estimate how many of these muons intersect the calorimeter volume.
We find that for $\nu_{\mu}$ ($\nu_e$) interactions, 7\% (48\%) of the events contain no muon, 48\% (34\%) contain one muon, 30\% (13\%) contain two muons, and 15\% (5\%) contain more than two mouns. The maximum muon energy is on average 38 GeV and 22 GeV for $\nu_{\mu}$ and $\nu_e$, respectively.
These results suggest that, while accompanying muons are common, their multiplicity and energy are not at a level that would overwhelm the calorimeter response. The remaining muons can be vetoed using standalone muon tracks reconstructed in the outer layers of CMS and ATLAS.
On the other hand, these muons could themselves provide valuable information: they offer a probe of hadron-induced muon multiplicities in this pseudorapidity region, and could also serve as an in-situ calibration tool for the associated neutrino flux, since both originate from the same parent hadrons.

However, we stress that this estimate is based on geometric propagation only and a more realistic assessment based on detailed simulations would be required to give a more precise estimation.
In addition, the impact of pileup from concurrent pp collisions must be evaluated to determine the feasibility of neutrino reconstruction in realistic LHC conditions.

Another important background for this configuration is the contamination of $\nu_e$ CC interactions by neutral-current (NC) neutrino interactions. 
In particular, the hadronic shower produced in an NC interaction can contain a hard $\pi^{0}$ whose electromagnetic shower may mimic the signature of an electron.

To estimate the size of this background, we evaluate the NC event yield in the CMS hadronic endcap calorimeter (HE). 
For the nominal configuration, we obtain approximately 3500 NC interactions, of which about 70\% contain at least one $\pi^{0}$.
Applying the same kinematic selections used in the analysis -- namely requiring the hadronic energy excluding the leading $\pi^{0}$ to be above 10 GeV and the energy of the leading $\pi^{0}$ to exceed 3 GeV -- we find that the surviving NC background is at a level comparable to the expected $\nu_e+\bar{\nu}_e$ signal contribution.

These numbers provide a first estimate of the achievable purity in the electron-neutrino channel. 
We stress, however, that this estimate is based only on particle-level information and does not include a realistic detector response.
In particular, the discrimination between electromagnetic showers initiated by electrons and those originating from $\pi^{0}\to\gamma\gamma$ decays is expected to depend sensitively on the calorimeter geometry, shower topology, and reconstruction performance for interactions initiated inside the calorimeter volume. 
Additional handles, such as the longitudinal shower development and energy-loss profile, could potentially be exploited to further suppress this background~\cite{MINERvA:2023ner}.
A dedicated detector-level simulation will therefore be required to obtain a reliable estimate of the NC contamination in the $\nu_e$ sample.

Finally, most of the opportunities described above will require reconstruction of the neutrino energy.
Useful neutrino energy resolutions of order 10--20\%, as achieved in several previous fixed-target neutrino experiments, will require adequate containment of the hadronic shower together with reliable reconstruction of the outgoing charged lepton. 
These requirements will likely restrict the analysis to a subset of the detector volume with sufficient containment and tracking capabilities, thereby reducing both the effective fiducial mass and the accessible pseudorapidity range. 
The event yields presented in this work should therefore be interpreted as optimistic upper limits for physics analysis, since they are based on the total instrumented mass of the relevant detector subsystems.

\section{Conclusions}

In this work, we have demonstrated that SHIFT@LHC, a proposed gaseous fixed target located 160 meters upstream of IP5 or IP1, would produce a measurable flux of neutrinos detectable in the CMS and ATLAS calorimeters.

Our estimates indicate that $\mathcal{O}(10^4)$ neutrino interactions could be observed using just 1\% of the LHC Run-4 integrated luminosity, with energies spanning from 20 GeV up to 1 TeV. 
The majority of events arise from $\nu_\mu$ CC interactions, but a non-negligible sample of $\nu_e$ events is also expected, allowing for flavor-dependent studies. 
These interactions occur predominantly in the hadronic calorimeter, though the electromagnetic calorimeter also contributes at a significant level.

Future work should refine these results with more detailed simulations of hadron propagation through the LHC tunnel, including the impact of local structures and surrounding material. 
Such studies could also identify optimal target locations to maximize neutrino yields. 
In addition, a realistic assessment of pileup and detector performance at CMS and ATLAS will be essential to establish the feasibility of neutrino reconstruction under standard LHC conditions.
In particular, dedicated reconstruction studies will be required to quantify the achievable flavor identification efficiency and neutrino energy resolution.

If realized, this setup would mark the first observation of neutrinos in one of the general-purpose detectors of LHC. 
Beyond its technical novelty, such a measurement would open a new window into neutrino production and interactions in a forward pseudorapidity regime that is relevant to atmospheric neutrino experiments.
By probing hadron production in collisions with light nuclei and neutrino cross sections in different materials, SHIFT could provide valuable input for predictions in experiments such as KM3NeT-ORCA, IceCube-Upgrade, DUNE, and Hyper-Kamiokande.

\textbf{\textit{Acknowledgments ---}} We thank Albert De Roeck and Felix Kling for useful discussions about LHC neutrinos, and Juliette Alimena for the discussion and a thorough review of the manuscript. AGS is supported by the CDEIGENT Grant No. CIDEIG/2023/20, by the MICIU/AEI grant PID2024-156285NB-C41, the SO project CEX2023-001292-S, and a 2024 Leonardo Grant from BBVA Foundation. The BBVA Foundation is not responsible for the opinions, comments, and contents included in the project or the results derived therefrom, which are the responsibility of the authors. JN would like to acknowledge the support from DESY (Hamburg, Germany), a member of the Helmholtz Association HGF, and support by the Deutsche Forschungsgemeinschaft (DFG, German Research Foundation) under Germany’s Excellence Strategy – EXC 2121 "Quantum Universe" – 390833306.

\bibliographystyle{apsrev4-2}
\bibliography{references}

\end{document}